\documentclass[12pt]{article}

\usepackage{newtxtext}   % Times New Roman-like text
\usepackage{newtxmath}   % Times New Roman-like math
\usepackage{titling}
\usepackage{siunitx}
\usepackage{multirow}

\pretitle{\centering\fontsize{20}{28}\selectfont}
\posttitle{\par}

\title{A Spatial Analysis of Proximity to the Built Environment and Wildfire Ignition Cause in the United States}

\usepackage{titlesec}

\titleformat{\section}
  {\normalfont\large\bfseries}
  {\thesection}{1em}{}

\titleformat{\subsection}
  {\normalfont\normalsize\bfseries}
  {\thesubsection}{1em}{}

\titleformat{\subsubsection}
  {\normalfont\small\bfseries}
  {\thesubsubsection}{1em}{}

\date{} % Leave the date empty to suppress it

\usepackage{graphicx} % Required for inserting images
\usepackage[backend=biber, style=authoryear, dashed=false, hyperref=true, doi=true, url=false, isbn=false, uniquename=false, uniquelist=false, sorting=nyt, maxcitenames=2, mincitenames=1, maxbibnames=50, minbibnames=5]{biblatex}

\usepackage[hidelinks]{hyperref}
\usepackage{xurl}
\DeclareFieldFormat{doi}{%
  \hfil\penalty90\hfilneg\space
  \ifhyperref
    {\href{https://doi.org/#1}{\nolinkurl{https://doi.org/#1}}}
    {\nolinkurl{https://doi.org/#1}}}

\renewbibmacro*{doi+eprint+url}{%
  \printfield{doi}}

\usepackage[letterpaper, left=1.0in, right=1in, top=1in, bottom=1in]{geometry}
\usepackage{amsmath}
\usepackage[dvipsnames]{xcolor}  % Color of text
\usepackage{spreadtab, xcolor, booktabs}
\usepackage{array}
\usepackage{caption}
\usepackage{parskip} % Handles paragraph spacing
\usepackage{soul}
\usepackage{enumitem}  % to creat list
\usepackage{tabularx}
\usepackage{float}

\usepackage{lineno}
\usepackage{etoolbox}

\AtBeginEnvironment{align}{\linenomath}
\AtEndEnvironment{align}{\endlinenomath}

\titlespacing*{\section}{0pt}{2.5ex plus 1ex minus .2ex}{1.6ex plus .2ex}
\titlespacing*{\subsection}{0pt}{2.3ex plus 1ex minus .2ex}{1.0ex plus .2ex}
\titlespacing*{\subsubsection}{0pt}{2.3ex plus 1ex minus .2ex}{1.0ex plus .2ex}

\begin{document}

\setlength{\droptitle}{-4.9em}
\maketitle

\vspace{-5.8em}

\begin{center}
{\large Ardavan Yazdanbakhsh}\\[0.3em]
{\small City College of New York, Department of Civil Engineering\\
CUNY Graduate Center, Earth and Environmental Sciences Program\\
New York, NY, United States\\
ayazdanbakhsh@ccny.cuny.edu}
\end{center}

\vspace{-0.1em}

% \linenumbers
% \modulolinenumbers[1]

        \section*{Abstract} 

\textbf{Background.} Human activities in areas with burnable land cover can lead to wildfire ignitions, and higher levels of human presence increase the likelihood of such fires. Building locations provide a widely available representation of human presence.  \\[0.35em]
\textbf{Aims.} This study examines how accurately proximity to buildings predicts wildfire ignition cause across different land-cover types.  \\[0.35em]
\textbf{Methods.} Supervised classification methods were applied to building, land-cover, and ignition data for the contiguous United States to estimate the conditional probability that an ignition is human-caused. Proximity to buildings was represented using either building density within surrounding zones or distance to nearby areas with high building density.  \\[0.35em]
\textbf{Key results.} Both methods, particularly the distance-based approach, showed that proximity to buildings is a strong predictor of ignition cause: approximately three quarters of both human- and lightning-caused ignitions were correctly classified. Retraining the models separately for individual states did not substantially improve classification performance relative to the national model.  \\[0.35em]
\textbf{Conclusions.} Building proximity is an effective predictor of wildfire ignition cause, and the relationship between ignition cause and building-related predictors is similar across many regions of the contiguous United States.  \\[0.35em]
\textbf{Implications.} Building-proximity variables can be incorporated into wildfire forecasting and risk assessment models to improve prediction of human-caused wildfire ignitions.

%         \section*{Abstract} 

% Human activities in areas with burnable land cover can lead to wildfire ignitions, and higher levels of human presence increase the likelihood of such fires. The spatial distribution of human presence is generally reflected by building locations, for which data are widely available. This study quantitatively assesses the effect of proximity to buildings on wildfire ignition cause across different land-cover types. Supervised classification methods were applied to building, land-cover, and ignition data for the contiguous United States to estimate the conditional probability that an ignition is human-caused. Two methods were used to represent proximity to buildings: one based on building density within surrounding zones and the other based on the distance to nearby areas with high building density. The findings from both methods, particularly the distance-based approach, showed that proximity to buildings is a strong predictor of ignition cause: approximately three quarters of both human- and lightning-caused ignitions were correctly classified. In addition, retraining the models separately for individual states did not substantially improve classification performance relative to the national model. This suggests that the relationship between ignition cause and the building-related predictors is similar across many regions of the contiguous United States.

\par\vspace{0.8em}
\textbf{Keywords:} Wildfire; Ignition cause; Building proximity; Human activity; Spatial classification.\par\vspace{0.5em}

        \section{Motivation and Objective}

The annual financial burden of wildfires in the United States is estimated to range between \$71.1 billion and \$347.8 billion \parencite{Thomas2017NIST1215}. Human ignitions are a major cause of wildfires in the United States. \textcite{Balch2017} showed that in the United States most wildfires are started by people, and human-caused fires extended the fire season and occurred under wetter fuel conditions and in areas with less lightning. Human influence on wildfire ignition has been examined in several studies. These studies model wildfire ignitions as a function of a number of explanatory variables, including measures of human influence, and compare the predictive performance of these variables.  

\begin{figure} [H]
    \centering
    \includegraphics[width=1.0\linewidth]{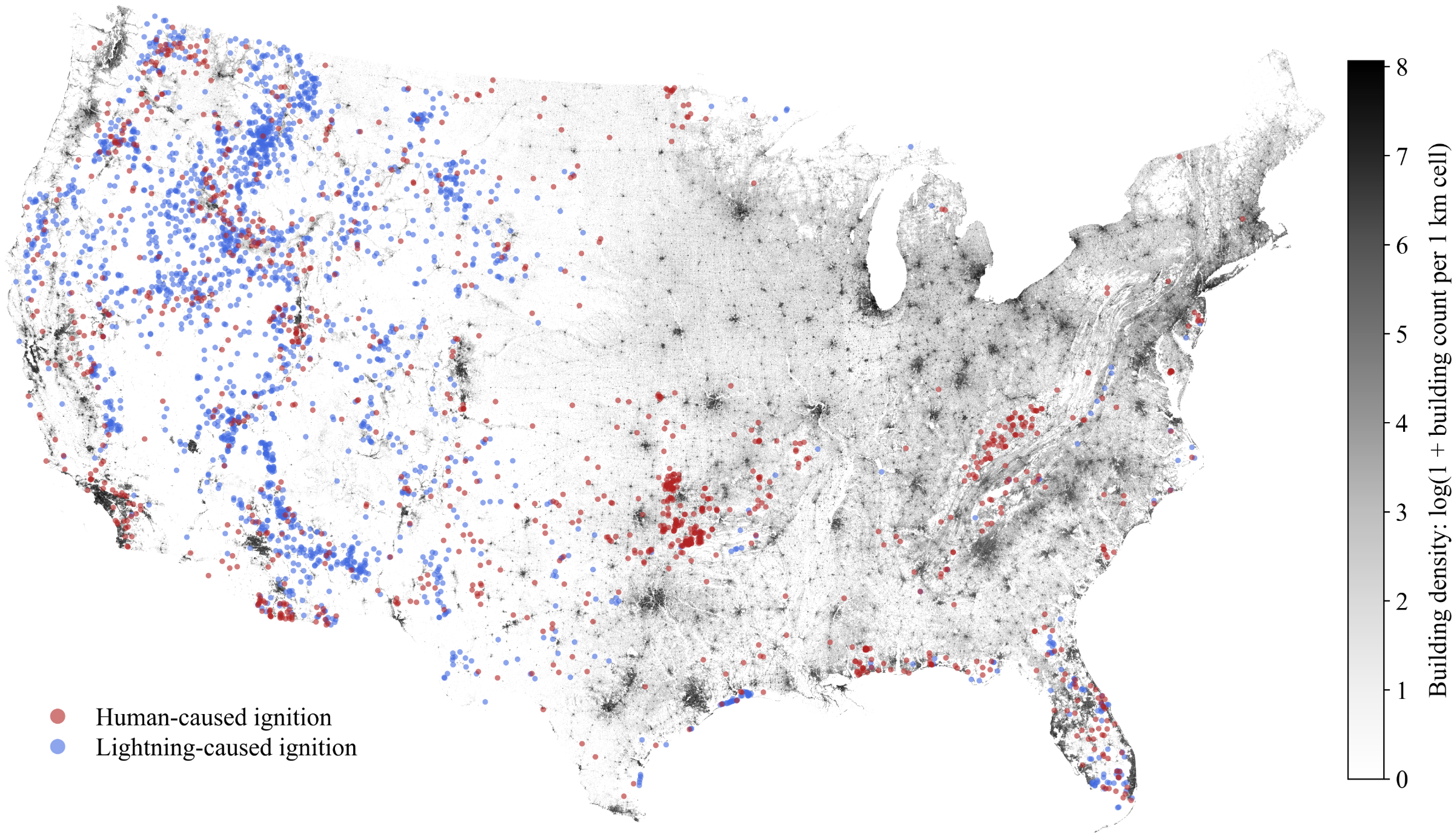}
    \caption{The wildfire ignition points used in this study are shown with distinct colors for each ignition type. Darker gray shades on the map indicate higher building density. The spatial distribution of the ignition points suggests that human-caused fires tend to occur closer to areas with higher building density. }
    \label{fig:BuidingDensity_IgnitionType}
\end{figure}

Most buildings threatened or destroyed by wildfires in the United States are located within the wildland–urban interface, where residential development intermingles with wildland vegetation \parencite{Kramer_2018}. This study examines quantitatively a related but distinct hypothesis: that wildlands in close proximity to buildings are more susceptible to human-caused wildfire ignitions. Some studies in the past two decades report patterns consistent with this hypothesis. For example, spatial analysis of ignition points in Portugal revealed that over 70\% of wildfires occurred in municipalities with a population density exceeding 100 inhabitants per square kilometer \parencite{catry2007}. Furthermore, a broad review of European fire drivers shows that the majority of these ignitions are concentrated within a 500-meter buffer of roads and urban interfaces \parencite{ganteaume2013}. Figure \ref{fig:BuidingDensity_IgnitionType} shows building density in the contiguous United States and the wildfire ignition points used in this study (Section \ref{sec:IgnitionData}). The spatial distribution of the ignition points reveals that human-caused ignitions are more likely to occur closer to areas with higher building density.

\vspace{-\parskip}

Predicting whether and when a wildfire may occur in a specific region requires the spatiotemporal analysis of environmental and meteorological data and remains an active area of research. The goal of this study is not to develop predictive models for wildfire occurrence. Instead, we aim to quantify how proximity to buildings influences the type of wildfire ignition. Specifically, given that a wildfire occurs at a location, we examine how well proximity to buildings can explain the conditional probability that the fire is human-caused. Although prior studies have used predictors such as population density and proximity to roads to represent human presence, this study investigates whether building location, for which data are widely available, has the potential to be a useful predictor. Understanding how a readily available variable such as proximity to buildings relates to wildfire ignition type has practical implications for prevention and response strategies. Human-caused fires can often be reduced through targeted policies and public awareness, whereas lightning-caused fires are less preventable and require emphasis on detection and suppression. Identifying where each ignition type is more likely to occur can help guide resource allocation, inform public education efforts \parencite{Naser2025}, and enhance risk management.

This study focuses on wildfires in the contiguous United States. The analysis is based on ignition data from 1999 to 2020, land cover information and spatial data on building locations. To integrate these datasets, we constructed a relatively fine 1 km by 1 km grid over the study region. Each grid cell was characterized by its land cover category, building density, and the presence or absence of wildfire ignition, along with the ignition type when applicable. A unique feature of this study is that, in addition to attributes of the focal cell, it also used measures of the built environment in surrounding cells. Based on this framework, we modeled the conditional probability that an ignition is human-caused given that a wildfire has occurred. Spatial cross-validation was used to assess the model predictive performance across regions. To highlight the contribution of this work, the next section reviews relevant studies on the relationship between indicators of human presence and wildfire ignition.

        \section{Background} \label{sec:Background}

\enlargethispage{\baselineskip}\textcite{Krawchuk2009} modeled the global distribution of wildfire occurrence at a coarse spatial resolution (approximately 100 km grid cells) using multiple temperature and precipitation variables, burnable biomass, lightning flash density, and the Human Footprint Index \parencite{Sanderson2002,Venter2016}. The Human Footprint Index is a composite measure of human pressure that combines population density with spatial data on land transformation such as croplands and settlements, and infrastructure such as roads, railways, navigable waterways, and nighttime lights. Their results showed that climate and fuel availability largely determine where fires can occur at broad spatial scales, and that human influence modifies fire probability within those regions. In fuel-rich and moderately populated landscapes, higher human influence was associated with increased fire probability due to additional human ignitions, whereas in densely populated and intensively managed landscapes fire probability declined because land conversion, infrastructure, and fire suppression measures reduce the continuity of burnable vegetation and limit fire spread.

\textcite{Calef2008} examined how wildfire occurrence in interior Alaska from 1988 to 2005 was related to proximity to human access routes and settlements. Their analysis focused on two outcomes: wildfire ignition frequency and the proportion of area burned. As predictors, they considered distance to settlements (represented as GIS point locations of towns and villages) and distance to major highways and navigable rivers. The results showed that ignition frequency was highest near settlements and access routes, whereas the proportion of area burned increased with distance from settlements, suggesting that fires near populated areas were more likely to be suppressed before spreading extensively.

\textcite{Hawbaker2013} evaluated the relative influence of human, vegetation, climate, and topographic variables on wildfire occurrence across the United States. Their models included several human-related predictors, such as housing density, road density, and proximity to roads and settlements, which represent the intensity of human presence and access in a landscape. Housing density was derived from housing-unit counts in the 2000 U.S. Census and therefore represents residential housing only, not commercial or industrial buildings. The results showed that the human variables were among the strongest predictors of wildfire occurrence. In particular, ignition hotspots were frequently located in areas with relatively high housing density within the wildland–urban interface and in regions surrounding large metropolitan areas.

At the global scale, \textcite{Bistinas2013} analyzed the relationship between human population density and burned area using fire data aggregated to 0.5° grid cells (about 50 km resolution). Their results showed that the relationship differs substantially across regions. In some areas, particularly rangelands, burned area increases with population density because human land-use practices that involve the use of fire, such as pasture management and vegetation clearing, promote burning. In addition, higher population density increases the likelihood of accidental ignitions. In other areas, especially croplands and densely settled landscapes, burned area decreases as population density increases because agriculture, infrastructure, and intensive land management break up continuous natural vegetation and fires are more actively suppressed. Their analysis was conducted at a relatively coarse spatial resolution and examined associations between population density and burned area within individual grid cells, without explicitly accounting for finer-scale spatial interactions or cross-cell effects.

At the continental scale, \textcite{Parisien2016} modeled wildfire probability across North America at a 1 km resolution for the period 1984–2014. The models related the probability that a given pixel burned at least once during the study period to climatic variables (e.g., temperature and precipitation), lightning density, and landscape characteristics such as topography and the proportion of nonfuel areas. Human influence was represented by population density, the Human Footprint Index, and a measure of remoteness based on `roadless' volume. The results showed that, after controlling for climatic and landscape factors, human influence was often an important predictor of wildfire probability. In many regions fire probability declined with increasing human influence, consistent with landscape fragmentation and fire suppression in developed areas, whereas in some remote regions fire probability increased with human presence, suggesting that fire occurrence there may be limited by ignition sources.

\textcite{Syphard2007} examined the influence of human settlement patterns on fire frequency and area burned in California using county-level analyses. Incorporating Census-based housing unit counts, road density, and proximity to the wildland–urban interface, they found that human factors were significant predictors of fire activity. Consistent with findings from other studies, the relationship was nonlinear, with fire activity increasing at intermediate levels of development and declining in highly urbanized areas. \textcite{Fusco2016} analyzed wildfire ignition locations across the western United States using satellite-derived burned-area data. They examined how the probability of ignition was associated with anthropogenic landscape features, including the extent of agricultural land cover, proximity to railroads, road networks, and other land-use disturbances. Their results showed that certain specific land-use features, particularly agricultural areas and railroad corridors, were stronger predictors of ignition occurrence than simple measures such as population or overall road density. They also found that the relative importance of these predictors differed across ecoregions.

\textcite{Gelabert2025} modeled human-caused wildfire ignition probability across five representative regions in Europe at 100 m spatial resolution. Ignition records were combined with gridded predictors, including distance to the wildland–urban interface, proximity to roads, and population density, which consistently ranked among the most important human-related variables, although their relative importance varied across regions. Comparing region-specific models with a single model fitted to all regions, the authors found that the latter model performed worse locally.

                \section{Study Design}

This section outlines the data sources, data preprocessing procedures, modeling framework, and validation approach used to analyze wildfire ignition causes. The models are trained and cross-validated both for the contiguous United States as a whole and separately for the 10 states with the largest numbers of ignition points in the study dataset. The section also describes the evaluation metrics used to assess predictive performance.

            \subsection{Data}

        \subsubsection{Ignition Types and Locations}  \label{sec:IgnitionData}

Wildfire ignition locations for the period 1999–2020 were obtained from the Incident Status Summary (ICS-209) wildfire incident dataset, a federal dataset maintained primarily by the U.S. Forest Service and containing operational reports prepared by incident management teams during major wildfire events \parencite{Denis2020}. The dataset includes reported point-of-origin coordinates for each incident as well as the incident cause, which was used to classify ignitions as human- and lightning-caused. In this study, we considered only large wildfires due to their disproportionate ecological and economic impacts \parencite{Balch2017,Nagy2018}. To identify ignition points corresponding to these fires, we used the Monitoring Trends in Burn Severity (MTBS) dataset, a U.S. federal dataset that provides georeferenced fire perimeters for large wildfires (at least 1,000 acres in the western United States and 500 acres in the eastern United States) \parencite{Short2017,Eidenshink2007}. We then retained ignition points from the ICS-209 dataset that fall within these perimeters. The study period (1999–2020) corresponds to the temporal overlap between the ICS–209 and MTBS datasets. Figure \ref{fig:BuidingDensity_IgnitionType} shows the ignition points used in this study. 

It should be noted that although most wildfire ignitions in the United States are human-caused, the majority of ignition points in this study correspond to lightning ignitions (1,140 human-caused and 2,135 lightning-caused fires). This is due to the fact that we studied only large fires, and lightning-caused fires account for a disproportionately large share of total burned area, as they often occur in remote areas and under conditions conducive to fire spread, allowing them to grow larger before containment \parencite{Balch2017}.

        \subsubsection{Land Cover Categories}

Vegetation and fuel conditions at wildfire ignition locations were characterized using the LANDFIRE Fire Behavior Fuel Model 40 (FBFM40) dataset for the conterminous United States. LANDFIRE is a U.S. federal dataset that provides spatially consistent information on surface fuel conditions relevant to wildfire behavior, in which each location is assigned a fuel model code corresponding to a specific fuel type \parencite{Rollins2009, ScottBurgan2005}. Because FBFM40 includes a large number of fuel model categories, using them individually would result in an excessive number of predictors and sparsity across categories, complicating model estimation and interpretation. Therefore, guided by ecological similarity, the fuel models were aggregated into five broader groups: Grassland, Shrub/Chaparral, Timber/Forest, Agriculture/Managed, and Developed. Each wildfire ignition location was assigned the fuel model corresponding to its geographic location.

Developed areas and agricultural or otherwise managed lands were retained as separate groups because ignition mechanisms in these environments differ from those in wildland fuels. Fuel models corresponding to water (92), snow/ice (98), and barren (99) were excluded because they do not represent meaningful burnable fuel conditions. The mapping between the aggregated categories and the underlying fuel models is provided in Table~\ref{tab:fuel_aggregation}. LANDFIRE data are released periodically, with major updates available for 2001, 2008, 2010, 2012, 2014, 2016, 2019, and 2022. Because fuel model classification and mapping can vary across releases, using the version closest in time to each ignition could introduce inconsistencies. Therefore, the 2019 LANDFIRE fuel model layer was used for all ignition locations.

\begin{table}[htbp]
\centering
\caption{Aggregation of fuel models into broader categories used in this study. Fuel model names follow the standard classification of \textcite{ScottBurgan2005}.}
\label{tab:fuel_aggregation}

\fontsize{11}{13}\selectfont
\renewcommand{\arraystretch}{1.15} % <-- increases row height (gap between rows)
\setlength{\extrarowheight}{2pt}  % <-- extra padding inside rows

\newcommand{\catw}{2.1cm} % Width of 1st column; adjust as needed

\begin{tabular*}{\textwidth}{@{\extracolsep{\fill}} p{\catw} p{\dimexpr\textwidth-\catw-2\tabcolsep\relax}@{}}
\hline
\noalign{\vskip 1pt}
Aggregated category & Fuel model (official code) \\
\noalign{\vskip 2.0pt}
\hline
\noalign{\vskip 2pt}

Grassland &
Short Grass (101); Timber Grass and Understory (102); Tall Grass (103); Grass--Shrub (202) \\[4pt]

Shrub/\newline Chaparral &
Shrub (104); Brush (105); Dormant Brush, Hardwood Slash (106); Chaparral, Low Load (121); Chaparral, Moderate Load (122); Chaparral, High Load (123); Shrub Litter (181); Timber--Shrub (182); Shrub--Grass (183); Slash--Shrub (184); Very Sparse Grass or Shrub (93) \\[4pt]

Timber/\newline Forest &
Closed Timber Litter (108); Short-Needle Timber Litter (141); Short-Needle Timber with Grass Understory (142); Short-Needle Timber with Shrub Understory (143); Deep Long-Needle Litter (144); Long-Needle Litter (145); Hardwood Litter (146); Southern Rough (147); Closed Timber Litter, Heavy Dead and Down (148); Hardwood Litter with Dead and Down Woody Material (149); Light Slash (161); Medium Slash (162); Heavy Slash (163); Short-Needle Litter with Shrub Understory (164); High Load Conifer Litter with Shrub Understory (165) \\[4pt]

Agriculture/\newline Managed &
Nonburnable Agricultural Land (185); Agricultural or Herbaceous Land with Woody Encroachment (186); Orchard or Vineyard (188) \\[4pt]

Developed &
Urban/Developed (91); Developed Area / Mixed Human-Dominated Cover (189) \\

\noalign{\vskip 2pt}
\hline
\end{tabular*}
\end{table}

            \subsubsection{Building Locations} \label{sec:BuildingData}
            
Building data were obtained from the Microsoft Building Footprints dataset, which provides polygon representations of individual building footprints across the United States \parencite{MicrosoftBuildingFootprints}. Each record includes the geometric outline of a building. The dataset does not include detailed structural attributes such as building height, number of floors, construction type, or occupancy classification. Nevertheless, it provides high-resolution, nationwide coverage of building locations, which is essential for the proximity-based analyses of wildfire ignitions in this study.

Several alternative sources of building data were considered, including parcel-based inventories and census-derived measures, but these are either not publicly available nationwide or are available only at aggregated spatial scales. It should be noted that the Microsoft dataset is compiled from imagery acquired at different times and updated periodically; therefore, it does not correspond to a specific year or to the timing of individual wildfire events. As a result, the building data should be interpreted as an approximation of the built environment rather than a time-specific record of structures present at the time of each fire.

        \subsection{Modeling Framework}  \label{sec:ModelFramwork}

In this study, two methods were used to model the relationship between ignition type and proximity to buildings across different land cover categories. Some of the components shared by both approaches are described first, followed by a detailed description of each method. Although the ignition data span multiple years, the analysis focuses on spatial rather than temporal variation. Ignition events are sparse in time at the grid-cell level, particularly because the analysis is restricted to large fires, making time series modeling at this resolution potentially unstable. For this reason, the data were aggregated across years and analyzed using a spatial framework. Temporal dynamics, including seasonal and interannual variation in ignition patterns, may provide additional insight and are an important direction for future research.

\enlargethispage{\baselineskip}In both methods, logistic regression is used to relate ignition cause to two sets of predictors: variables describing proximity to buildings and those representing land cover and fuel type. The building-related predictors distinguish the two methods. The ring-based method uses building densities within concentric rings surrounding the ignition cell, whereas the distance-based method uses the minimum distance from the ignition cell to areas exceeding specified building-density thresholds. The land-cover-related variables are identical in both methods and consist of the five categories listed in Table \ref{tab:fuel_aggregation}. These variables are treated as categorical predictors, with Timber/Forest used as the reference category in both analyses. The two methods are conceptually related to established approaches for characterizing spatial neighborhood effects. The ring-based method is analogous to moving-window or buffer-based analyses, in which summary measures (e.g., counts or densities) are computed within a set of surrounding zones at increasing distances from a focal location \parencite{OSullivan2010}. The distance-based method is related to distance-decay formulations, which assume that the influence of a feature decreases with distance, a principle widely used in spatial interaction and accessibility modeling \parencite{Tobler1970}.

To integrate wildfire ignition, fuel, and building data, the study area was divided into a regular grid of 1 km by 1 km cells, which served as the unit of analysis. For each cell, wildfire ignition information was assigned based on the presence of an ignition point within the cell, with separate indicators for human- and lightning-caused ignitions. One of the five fuel-type categories in Table \ref{tab:fuel_aggregation} was assigned to each cell. Because a cell may contain more than one fuel type, the predominant category was determined by evaluating five locations within the cell (the center and four interior points) and selecting the most frequently occurring fuel type. Building information was incorporated by counting the number of buildings within each cell, resulting in a measure of building density at the cell level. As noted in Section \ref{sec:BuildingData}, the dataset does not include building type, height, or occupancy. Therefore, this measure of building density is approximate. Each cell was thus characterized by a single fuel-type category, a building count, and ignition indicators.

\begin{figure}[H]
\centering
\includegraphics[width=0.5\textwidth]{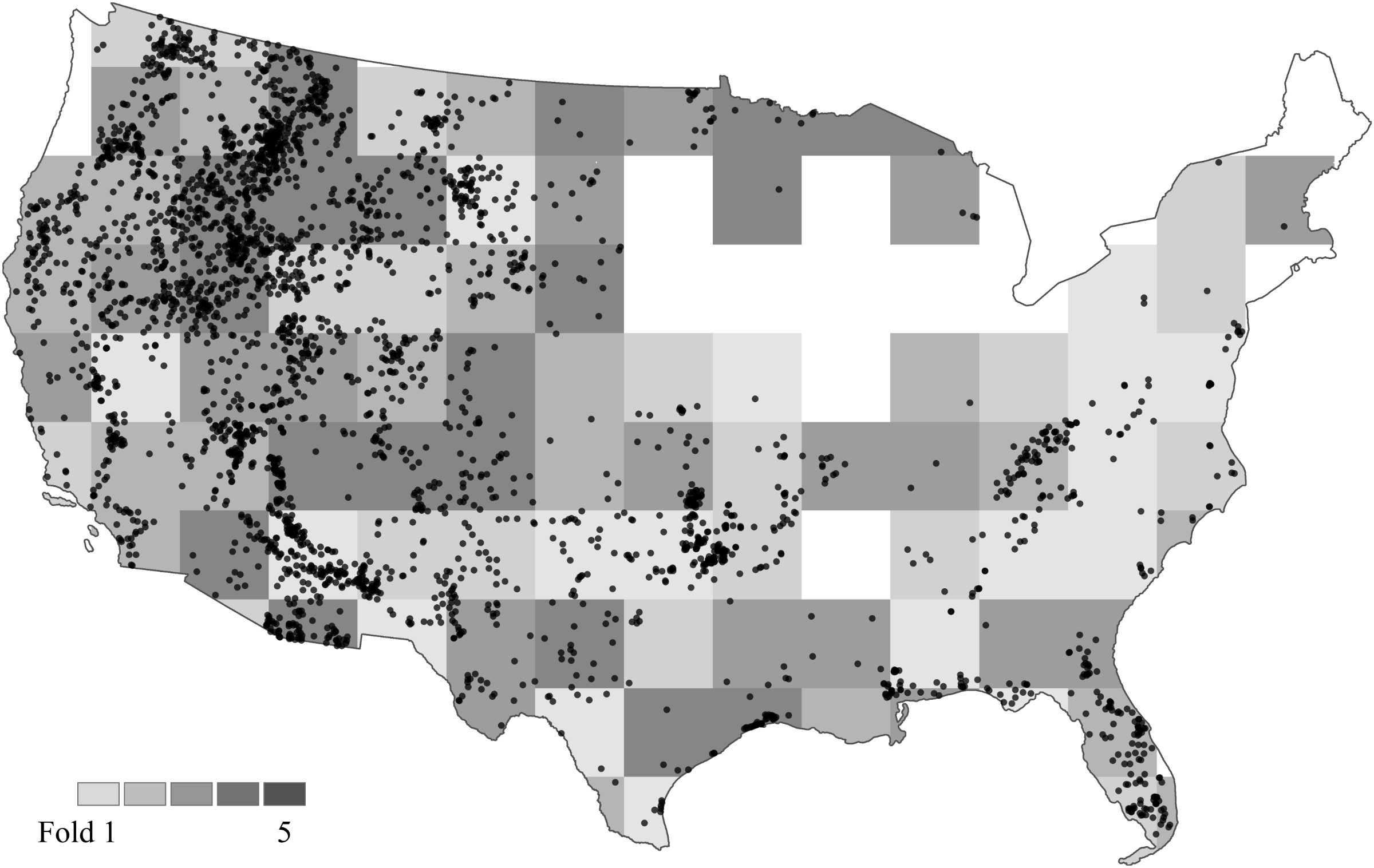}
\caption{For 5-fold cross-validation, the conterminous United States was partitioned into approximately 300 km × 300 km tiles, which were assigned to folds such that each fold contained a comparable number of ignitions.}
\label{fig:tiles_cv}
\end{figure}

To evaluate model performance in both methods, we used a spatial 5-fold cross-validation strategy \parencite{Hastie2009,Roberts2017}. For the national analysis, the study region was partitioned into approximately 300 km × 300 km non-overlapping tiles, resulting in 148 spatial units across the conterminous United States. All grid cells within each tile were assigned to the same fold to maintain spatial separation between training and test data. Tiles were allocated to folds using a greedy balancing procedure: they were first sorted in descending order based on the number of ignition events they contained, and then sequentially assigned to the fold with the fewest ignition points. This approach produces folds with similar numbers of ignition events while preserving spatial separation. The resulting spatial distribution of the five folds is shown in Figure~\ref{fig:tiles_cv}. For the state-level analyses, the same 5-fold spatial cross-validation framework was used separately within each state. To maintain a comparable number of spatial units across analyses, smaller non-overlapping tiles of approximately 75 km × 75 km were used for the state-level partitioning.

\vspace{0.4em}

    \subsubsection{Ring-Based Analysis}

In this method, proximity to buildings was quantified using building densities in 1 km–thick rings surrounding each cell with at least one ignition point. These square-shaped rings were defined using expanding square windows (1×1, 3×3, 5×5, 7×7, and 9×9) centered on each cell, with each ring corresponding to the area between successive windows. The correlations among the ring-based variables were high. In particular, correlations between building densities in adjacent rings exceed 0.84 and reach up to 0.95 for the outer rings (Table~\ref{tab:ring_corr}). This indicates substantial redundancy, suggesting that including multiple ring-based measures provides limited additional information for the analysis. We therefore used only the building density in each cell and in its outermost ring (ring 4), as two separate predictors.

\begin{table}[htbp]
\centering
\caption{Correlation matrix of log-transformed building densities for the center cell and four surrounding rings.}
\label{tab:ring_corr}

\fontsize{10}{12}\selectfont
\setlength{\tabcolsep}{4pt}
\renewcommand{\arraystretch}{1.0}   % keep rows tight
\setlength{\extrarowheight}{1pt}    % minimal padding

\begin{tabular}{lccccc}
\hline
\noalign{\vskip 1pt}
 & Center & Ring 1 & Ring 2 & Ring 3 & Ring 4 \\
\noalign{\vskip 2pt}
\hline
\noalign{\vskip 2pt}

Center  & 1.00 & 0.84 & 0.78 & 0.74 & 0.72 \\
Ring 1  &      & 1.00 & 0.93 & 0.88 & 0.84 \\
Ring 2  &      &      & 1.00 & 0.94 & 0.90 \\
Ring 3  &      &      &      & 1.00 & 0.95 \\
Ring 4  &      &      &      &      & 1.00 \\

\noalign{\vskip 2pt}
\hline
\end{tabular}
\end{table}

Since the goal of the study is to model the conditional probability of a human-caused fire given that a wildfire has occurred, the analysis in both methods was restricted to grid cells with an ignition. The dependent variable was defined as an indicator of ignition type, taking the value 1 for human-caused fires and 0 for lightning-caused fires. The independent variables included the logarithm of building density in the cell and in its outermost ring (ring 4), along with categorical variables representing the fuel type. The logarithmic transformation was used to reduce the influence of very large values. As mentioned earlier, a logistic regression model was used to estimate this conditional probability.

    \subsubsection{Distance-Based Analysis}

In the ring-based analysis, a ring may contain one or a few zones with high building concentrations while having a low overall building density. In such cases, the ring may not reflect the proximity of the corresponding ignition point to those zones. This limitation motivated the development of a method that measures proximity to buildings based on the distance to high-density grid cells. In this approach, for each cell with an ignition, we computed the distance to the nearest grid cell whose building count exceeded a specified threshold. Three such variables were constructed, corresponding to thresholds of 10, 100, and 500 buildings. These thresholds were selected after exploratory analysis to represent different levels of building density while avoiding highly correlated predictors. As in the first method, the analysis was restricted to grid cells with an ignition, and the dependent variable was an indicator of ignition type. The independent variables included the logarithm of building density in each cell, the distance-based measures described above, and categorical variables representing the fuel type of each cell. The distance variables were standardized to have zero mean and unit variance to place them on a comparable scale. As in the ring-based method, a logistic regression model was used to estimate the conditional probability that an ignition is human-caused given that a fire has occurred.

    \subsubsection{Model Evaluation Metrics}

To evaluate model performance, several complementary metrics were used, each representing a different aspect of how well the model distinguishes between human- and lightning-caused ignitions. This study is formulated as a supervised binary classification problem, in which ignition points are assigned to human- or lightning-caused categories based on explanatory variables. Accordingly, the evaluation relied on metrics commonly used in classification settings to assess discriminative ability and predictive performance \parencite{FAWCETT2006}. We treat human-caused ignitions as the positive class and lightning-caused ignitions as the negative class for modeling purposes; this convention is adopted solely for defining classification metrics and does not imply greater importance of one ignition type over the other. A brief description of the metrics and their interpretation in the context of this study is provided below.

Sensitivity, or true positive rate, is the proportion of human-caused ignitions that are correctly classified as human-caused. It is therefore a measure of how well the model identifies human-caused fires. Specificity, or true negative rate, is the proportion of lightning-caused ignitions that are correctly classified as lightning-caused, and measures how well the model identifies lightning-caused fires. Precision is the proportion of ignitions classified as human-caused by the model that are truly human-caused, and represents the reliability of positive predictions. Similarly, negative predictive value (NPV) is the proportion of ignitions classified as lightning-caused by the model that are truly lightning-caused, and represents the reliability of negative predictions. A classification threshold that yields relatively high values for both sensitivity and specificity indicates that the model is able to identify both human- and lightning-caused fires with good accuracy, without strongly favoring one class over the other. It should be noted that, given the proportion of human-caused fires in the data, precision and negative predictive value can be derived from sensitivity and specificity. Nevertheless, these metrics provide additional insight into the reliability of the model’s predictions for each class.

A receiver operating characteristic (ROC) curve is a standard tool in binary classification that illustrates the ability of a model to distinguish between two classes across different classification thresholds. It is constructed by plotting the true positive rate (sensitivity) against the false positive rate, which is equal to one minus specificity, as the decision threshold varies. A model with good discriminative ability produces a curve that lies closer to the upper-left corner of the plot, indicating high sensitivity and a low false positive rate. The area under the curve (AUC) summarizes this performance: values closer to 1 indicate better discriminative ability, whereas a value of 0.5 corresponds to random guessing. The AUC is particularly useful because it provides a comprehensive and threshold-independent measure of the model’s discriminative performance. Balanced accuracy at a given classification threshold is defined as the average of the model’s sensitivity and specificity at that threshold. It therefore represents the overall balanced performance of the model and, when both classes are equally important, can be used to help identify a threshold that provides an appropriate trade-off between the two.

            \section{Results and Discussion}

\begin{figure}
    \centering
    \includegraphics[width=1\linewidth]{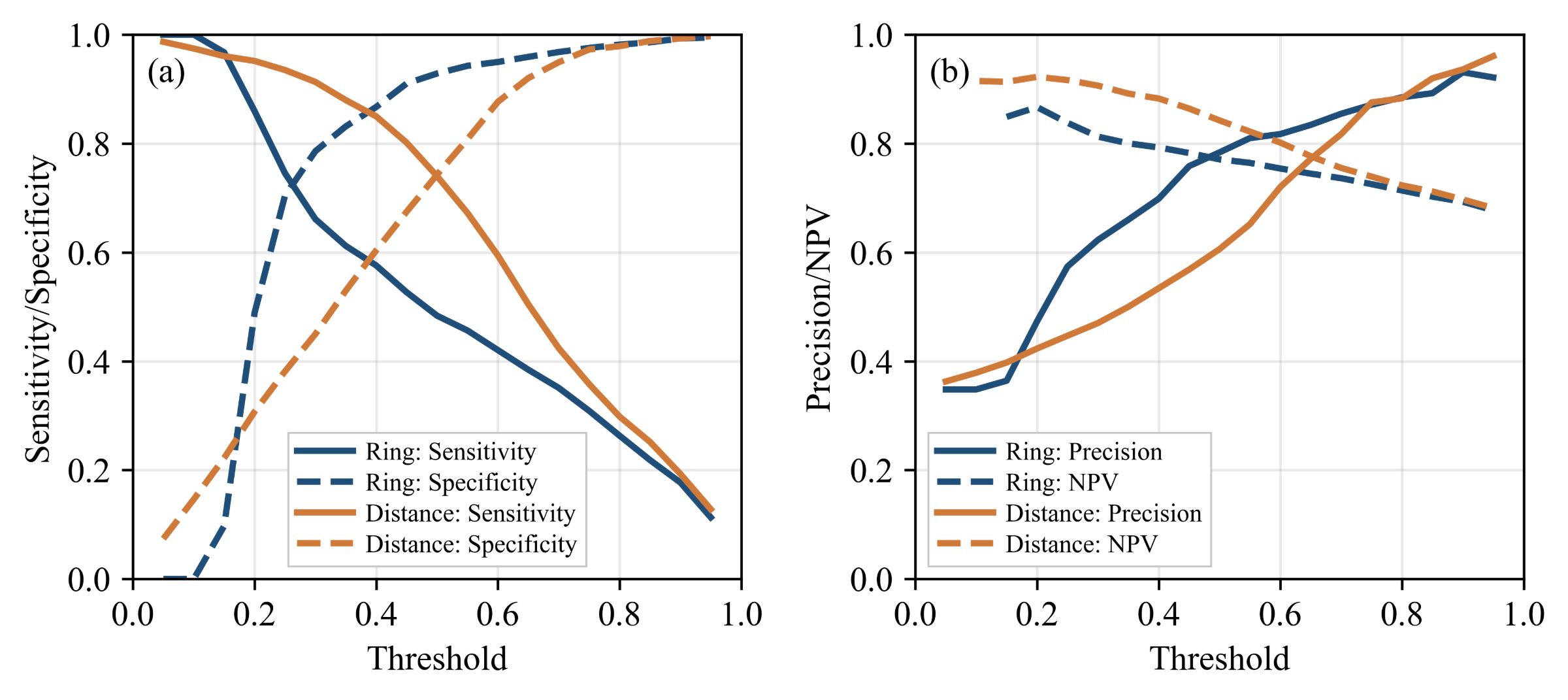}
    \caption{(a) Sensitivity and specificity across classification thresholds for the ring- and distance-based methods. (b) Precision and NPV across classification thresholds for the two methods.}
    \label{fig:The_Four_Metrics}
\end{figure}

Figure \ref{fig:The_Four_Metrics}.a presents the sensitivity and specificity curves across classification thresholds for both the ring- and distance-based models. The curves shows the expected trade-off between sensitivity and specificity as the classification threshold varies. At low thresholds, the models have high sensitivity but very low specificity, indicating that most human-caused ignitions are correctly classified, while many lightning-caused ignitions are incorrectly classified as human-caused; in this regime, the models are more likely to classify an ignition as human-caused. As the threshold increases, this pattern is reversed, and the models become more likely to classify an ignition as lightning-caused.

The ring- and distance-based models achieve balanced performance at thresholds of approximately 0.26 and 0.50, respectively, where sensitivity and specificity are equal. At these points, both measures are 0.72 for the ring-based model and 0.74 for the distance-based model. Therefore, at a threshold of 0.50, the distance-based method correctly identifies nearly three quarters of ignitions for both classes. It can be observed that, as the threshold moves away from the point at which sensitivity and specificity are approximately equal, both measures, particularly specificity, decline more rapidly in the ring-based model.

Figure \ref{fig:The_Four_Metrics}.b presents the precision and NPV curves at varying classification thresholds for both the ring- and distance-based models. The figure shows that, for both models, the threshold at which precision and NPV intersect differs from the threshold at which sensitivity and specificity are approximately equal. This indicates that no single threshold yields a simultaneous balance across all four metrics. At a threshold of 0.5, where sensitivity and specificity are both 0.74 for the distance-based model, the NPV is relatively high (about 0.84), meaning that 84\% of ignitions classified as lightning-caused are correctly identified. However, at the same threshold, precision is lower (about 0.61), indicating that only 61\% of ignitions classified as human-caused are truly human-caused. Whether this threshold is appropriate depends on the objective of the classification. If the goal is to detect as many human-caused fires as possible, and misclassifying a non-negligible number of lightning-caused fires as human-caused is acceptable, then thresholds that yield higher sensitivity at the expense of precision may be preferred.

For the distance-based model, a threshold of 0.65 yields the same value for precision and NPV (0.77). At this threshold, specificity is high (0.92), whereas sensitivity is relatively low (0.50), indicating that only 50\% of human-caused ignitions are correctly identified. Selecting a threshold that achieves high precision at the expense of sensitivity may be undesirable, as it results in many human-caused fires being misclassified as lightning-caused. As shown in Figures \ref{fig:The_Four_Metrics}.a and \ref{fig:The_Four_Metrics}.b, increasing the threshold improves precision and specificity but reduces sensitivity, while decreasing the threshold has the opposite effect. This trade-off is a fundamental characteristic of classification problems \parencite{Verbakel2020}.

\begin{figure}
    \centering
    \includegraphics[width=1\linewidth]{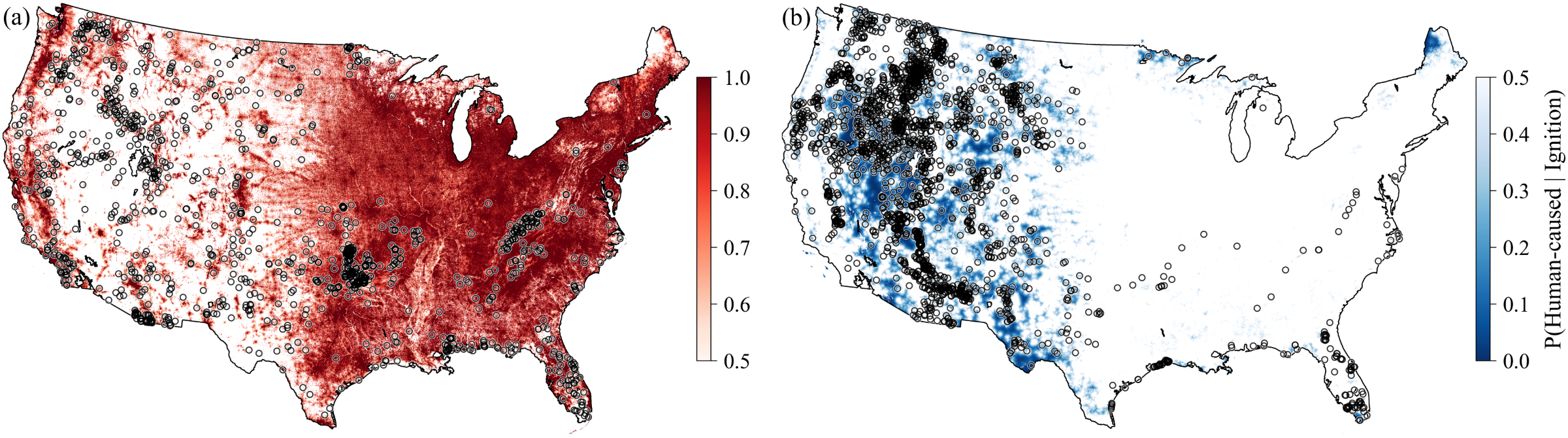}
    \caption{Risk maps of the contiguous United States based on the conditional probability of an ignition being human-caused given an ignition, created using the distance-based model. Map (a) shows regions where this probability is 0.5 or higher, along with the human-caused ignition points, and map (b) shows regions where the probability is below 0.5, along with the lightning-caused ignition points.}
    \label{fig:RiskMap_US}
\end{figure}

Based on the distance-based model, Figure \ref{fig:RiskMap_US} shows, in two separate maps, regions where the conditional probability of a human-caused ignition is above and below the threshold of 0.5. Specifically, Figure \ref{fig:RiskMap_US}(a) shows regions corresponding to probabilities higher than 0.5, along with the locations of the human-caused ignition points, and Figure \ref{fig:RiskMap_US}(b) shows regions corresponding to probabilities lower than 0.5, together with the lightning-caused ignition points. The figure shows that a large majority of the ignition points are located within regions corresponding to their correctly classified ignition type, as expected from the sensitivity and specificity results at this threshold. The figure reveals another important finding: regions classified as lightning-caused are generally located relatively close to at least one lightning-caused ignition point, whereas there are vast regions classified as human-caused that are far from any ignition points, whether human-caused or lightning-caused. One possible explanation is that most of the ignition points used in this study were lightning-caused and concentrated in specific regions, particularly in the West and Southeast (Figure \ref{fig:BuidingDensity_IgnitionType}). As a result, a spatial model is more likely to classify areas in those regions as lightning-caused. The human-caused ignition points, on the other hand, are more spatially dispersed. Consequently, regions with fewer ignition points are often closer to, or less far from, human-caused ignition points and are therefore more likely to be classified as human-caused. A larger dataset with more ignition points in those regions could affect the classification results.

\begin{figure}
    \centering
    \includegraphics[width=1\linewidth]{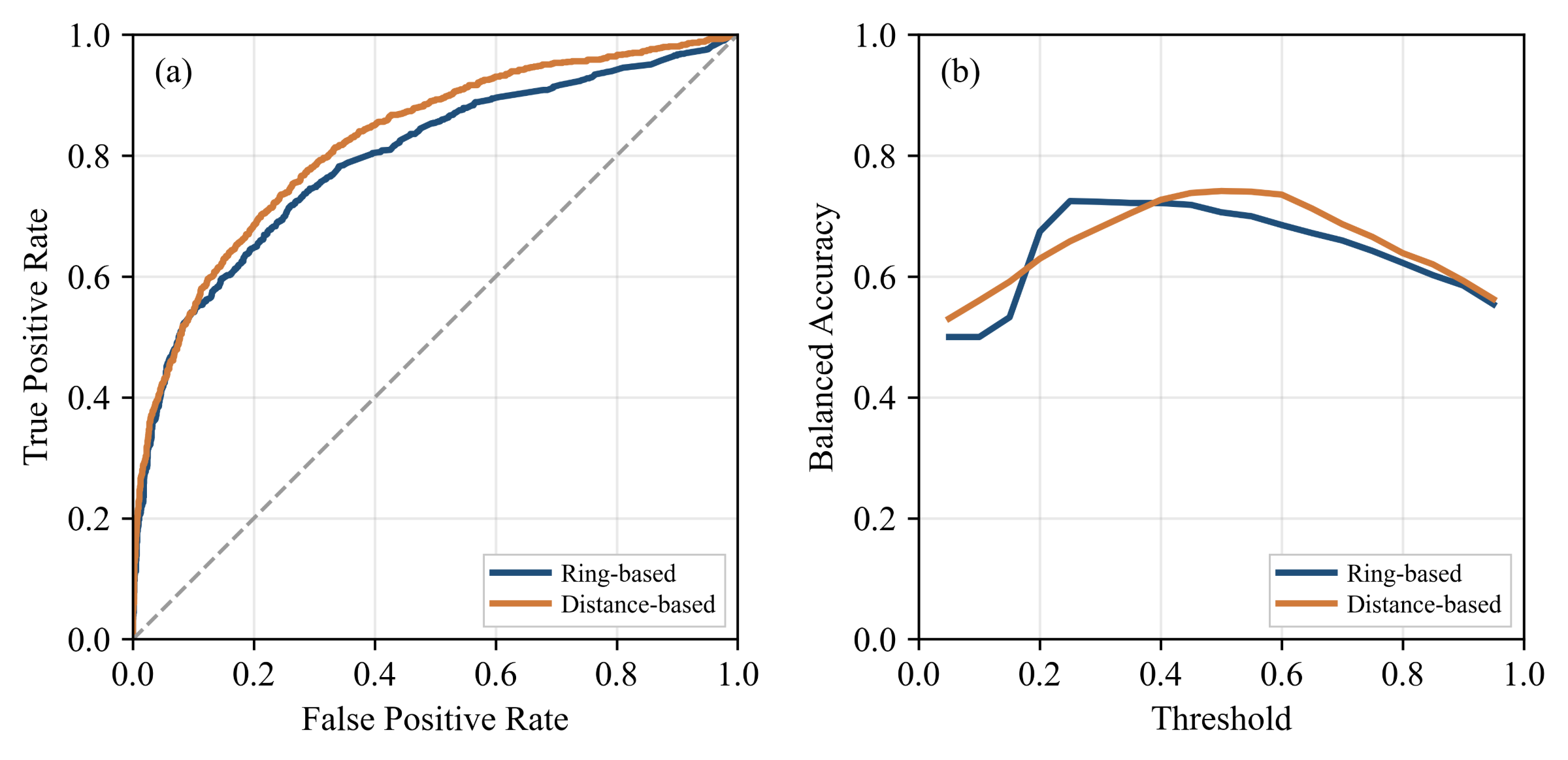}
    \caption{(a) ROC curves for the ring-based and distance-based methods; the corresponding AUC values are 0.79 and 0.82, respectively. (b) Balanced accuracy across classification thresholds for the two methods.}
    \label{fig:ROC_and_Balanced}
\end{figure}

Figure \ref{fig:ROC_and_Balanced}.a compares the receiver operating characteristic (ROC) curves for the ring- and distance-based models. The curve corresponding to the distance-based model lies slightly closer to the upper-left corner of the plot, indicating better discriminative ability across classification thresholds. This is consistent with the cross-validation results, where the distance-based model has a slightly higher mean AUC (0.818) than the ring-based model (0.797). A similar pattern is observed when comparing balanced accuracy across thresholds (Figure \ref{fig:ROC_and_Balanced}.b): the distance-based model has a slightly higher peak value (0.74 versus 0.73) and maintains comparable or higher performance over most thresholds, although the ring-based model performs slightly better within a limited interval (0.18 to 0.40). Overall, both models exhibit good discriminative performance, indicating that information on building locations and fuel type provides a useful basis for distinguishing between human- and lightning-caused ignitions.

Table \ref{tab:model_coefficients} presents the estimated coefficients for the ring- and distance-based models. The Timber/Forest category is not included in the table because, as discussed in Section \ref{sec:ModelFramwork}, it was used as the reference land-cover category in both models. Therefore, the coefficients reported for the remaining four land-cover categories indicate whether ignitions in those categories are more or less likely to be human-caused compared to ignitions in Timber/Forest areas. Positive coefficients indicate a higher likelihood of human-caused ignition relative to Timber/Forest, whereas negative coefficients would indicate a lower likelihood. Larger coefficient values indicate stronger differences from the Timber/Forest category.

The table shows that in both models, building-related variables are strongly associated with the probability of human-caused ignition. In the ring-based model, higher building density in both the focal cell and the surrounding ring is associated with an increased likelihood of human-caused fires, with coefficients of 1.26 and 0.96, respectively. In the distance-based model, building density remains positively associated with human-caused ignition, while distances to dense building areas are negatively associated, indicating that ignitions occurring farther from such areas are less likely to be human-caused. Fuel-related variables show generally consistent patterns across the two models. Relative to Timber/Forest areas, Developed and Agriculture/Managed areas have substantially higher probabilities of human-caused ignition, whereas Grassland and Shrub/Chaparral areas show smaller positive differences. The similarity of the coefficient patterns across the two modeling approaches suggests that the observed relationships are not strongly dependent on the specific representation of building proximity used in the analysis.

\begin{table}[htbp]
\centering
\fontsize{10}{12}\selectfont
\renewcommand{\arraystretch}{1.1}

\setlength{\heavyrulewidth}{0.5pt}
\setlength{\lightrulewidth}{0.5pt}
\setlength{\cmidrulewidth}{0.5pt}

\caption{Estimated coefficients of the final national ring-based and distance-based logistic regression models. Timber/Forest was used as the reference land-cover category in both models.}
\label{tab:model_coefficients}

% \begin{tabular}{llcc}
\begin{tabular}{p{3.3cm}lcc}
\toprule
Group & Predictor & Ring-based & Distance-based \\
\midrule

\multirow{5}{*}{Building variables}
& Log building density, focal cell          & 1.26 & 1.24 \\
& Log building density, outer ring          & 0.96 & --- \\
& Distance to cells with $\geq$10 buildings  & --- & -0.56 \\
& Distance to cells with $\geq$100 buildings & --- & -0.55 \\
& Distance to cells with $\geq$500 buildings & --- & -0.18 \\

\midrule

% \multirow{4}{*}{Land-cover categorical variables}
% \shortstack[l]{Land-cover\\[2pt]categorical variables}
\multirow{4}{*}{\shortstack[l]{Land-cover\\[2pt]categorical variables}}
& Grassland              & 0.68 & 0.57 \\
& Shrub/Chaparral        & 0.29 & 0.29 \\
& Agriculture/Managed    & 1.42 & 1.23 \\
& Developed              & 2.58 & 2.19 \\

\bottomrule
\end{tabular}

\end{table}

Table \ref{tab:state_model_comparison} presents model performance for the 10 states with the highest numbers of ignition points in the study dataset, measured by the sensitivity at the threshold where sensitivity and specificity are approximately equal. The table compares results obtained using the national model calibrated on the full dataset with those obtained using models calibrated separately for each state. Overall, the state-specific models did not produce substantial improvements over the national model. One possible reason is that the relationship between ignition cause and the predictors used in this study may be broadly similar across many states. For example, in many regions, human-caused ignitions tend to occur closer to developed areas and higher building densities, whereas lightning-caused ignitions are more common in more remote locations. If these relationships are similar across states, then a model trained using ignition data from the entire CONUS may already capture much of the relevant structure, leaving limited room for improvement from state-specific retraining. Another important factor is sample size. The national model was trained using all ignition points simultaneously, whereas each state-specific model used only a much smaller subset of observations. This reduction in training data may limit the ability of state-specific models to estimate relationships reliably, particularly for states with relatively few ignitions or highly imbalanced human- and lightning-caused ignition counts.

\begin{table}[htbp]
\centering
\fontsize{10}{12}\selectfont
\renewcommand{\arraystretch}{1.1}
\setlength{\heavyrulewidth}{0.5pt}
\setlength{\lightrulewidth}{0.5pt}
\setlength{\cmidrulewidth}{0.5pt}
\caption{Comparison of national and state-specific model performance using the distance-based method for the ten states with the highest numbers of wildfire ignitions in the study dataset, measured by the sensitivity at the threshold where sensitivity and specificity are approximately equal.}
\label{tab:state_model_comparison}

\begin{tabular}{
c
c
S[table-format=1.2]
S[table-format=1.2]
S[table-format=1.2]
S[table-format=1.2]
S[table-format=1.2]
S[table-format=-1.2]
}
\toprule
State &
Ignitions &
\multicolumn{1}{c}{\shortstack{Human\\fraction}} &
\multicolumn{1}{c}{\shortstack{National\\threshold}} &
\multicolumn{1}{c}{\shortstack{National\\sensitivity}} &
\multicolumn{1}{c}{\shortstack{State\\threshold}} &
\multicolumn{1}{c}{\shortstack{State\\sensitivity}} &
\multicolumn{1}{c}{Difference} \\
\midrule
ID & 428 & 0.18 & 0.47 & 0.74 & 0.22 & 0.77 & 0.03 \\
AZ & 327 & 0.23 & 0.47 & 0.59 & 0.24 & 0.52 & -0.07 \\
MT & 289 & 0.15 & 0.44 & 0.68 & 0.16 & 0.70 & 0.02 \\
OR & 240 & 0.15 & 0.39 & 0.64 & 0.16 & 0.67 & 0.03 \\
NV & 239 & 0.17 & 0.23 & 0.60 & 0.16 & 0.55 & -0.05 \\
CA & 218 & 0.37 & 0.54 & 0.79 & 0.33 & 0.81 & 0.03 \\
UT & 198 & 0.26 & 0.44 & 0.69 & 0.25 & 0.65 & -0.04 \\
OK & 196 & 0.97 & 0.70 & 0.80 & 0.98 & 0.62 & -0.18 \\
NM & 184 & 0.29 & 0.33 & 0.70 & 0.30 & 0.66 & -0.04 \\
FL & 131 & 0.52 & 0.56 & 0.54 & 0.49 & 0.54 & 0.00 \\
\bottomrule
\end{tabular}

\end{table}

The results of this study show that building proximity is a strong predictor for distinguishing between human- and lightning-caused ignitions, even within a modeling framework that relies on a limited set of static spatial predictors. Predictive performance can likely be further improved by incorporating additional data sources and modeling approaches. Because the present analysis was conducted using a subset of ignition points, the available sample size was limited. Expanding the dataset and incorporating additional predictors—such as proximity to other types of structures and infrastructure, weather conditions, and temporal patterns of human activity—as well as using alternative statistical or machine learning models, may improve predictive accuracy. Finally, while the present study focuses on spatial predictors, extending the framework to a spatiotemporal setting using a larger dataset with finer temporal resolution could provide a more comprehensive characterization of wildfire ignition causes.

            \section{Concluding Remarks}

This study quantitatively examined the relationship between proximity to buildings and wildfire ignition cause. For this purpose, classification models were developed using wildfire ignition points, land-cover data, and building-location data on a fine spatial grid together with measures that incorporate relationships among neighboring grid cells. The findings showed that variables derived from relatively simple and widely available data on the built environment can be used to distinguish between ignition causes. Despite several data and modeling limitations, including the use of a reduced set of ignition points and building and land-cover data that do not necessarily correspond to the timing of individual fires, the results show that building-proximity predictors can lead to relatively high classification accuracy. The comparison between models trained using national and state-level data suggests that the relationship between building proximity and wildfire ignition cause is similar in many regions of the contiguous United States. Wildfire forecasting models typically rely on a wide range of environmental and meteorological predictors. The findings of this work suggest that building-proximity variables may improve the predictive performance of such models, a hypothesis that should be examined in future studies.

            \printbibliography[heading=bibnumbered]

\vspace{0.5em}
\section{Statements and Declarations}

\textbf{Data Availability.} The data used in this study are publicly available from the sources identified and cited in the manuscript. No proprietary datasets were used. \\[0.35em]
\textbf{Conflict of Interest.} The author declares no conflicts of interest.

\end{document}